\documentclass[mathleft
]{an}
\usepackage{graphicx}
\usepackage{times}
\begin{document}

\Pagespan{789}{}
\Yearpublication{2006}%
\Yearsubmission{2005}%
\Month{11}%
\Volume{999}%
\Issue{88}%

\title{The PLATO End-to-End CCD Simulator -- \\
        Modelling space-based ultra-high precision CCD photometry for the assessment study of the PLATO Mission}

\author{W. Zima\inst{1}\fnmsep\thanks{Corresponding author:
  \email{zima@ster.kuleuven.be}\newline}
\and  T. Arentoft\inst{2}
\and  J. De Ridder\inst{1}
\and  S. Salmon\inst{3}
\and  C. Catala\inst{4}
\and  H. Kjeldsen\inst{2}
\and  C. Aerts\inst{1,5}
}
\titlerunning{The PLATO End-to-End CCD Simulator}
\authorrunning{W. Zima et al.}
\institute{
Instituut voor Sterrenkunde, K.U. Leuven, Celestijnenlaan 200D, 3001 Leuven, Belgium
\and 
Instituut for Fysik og Astronomi, Aarhus Universitet, Ny Munkegade Bygn. 520, DK-8000 \AA rhus C, Danmark
\and 
Institut d' Astrophysique et de G\'eophysique de l' Universit\'e de Li\`ege, All\'ee du 6 Ao\^ut, 17 B-4000 Li\`ege, Belgium
\and
LESIA, UMR8109, Universit\'e Pierre et Marie Curie, Universit\'e Denis Diderot, Observatoire de Paris, 92195 Meudon, France
\and
Department of Astrophysics, IMAP, University of Nijmegen, PO Box 9010, 6500 GL Nijmegen, The Netherlands
}

\received{30 May 2010}
\accepted{11 Nov 2010}
\publonline{later}


\abstract{
 The PLATO satellite mission project is a next generation ESA Cosmic Vision satellite project dedicated to the
 detection of exo-planets and to asteroseismology of their host-stars using ultra-high precision photometry. The main goal of the PLATO mission is to provide a full statistical analysis of exo-planetary systems around stars that are bright and close enough for detailed follow-up studies. Many aspects concerning the design trade-off of a space-based instrument and its performance can best be tackled through realistic simulations of the expected observations. The complex interplay of various noise sources in the course of the observations made such simulations an indispensable part of the assessment study of the PLATO Payload Consortium. We created an end-to-end CCD simulation software-tool, dubbed PLATOSim, which simulates photometric time-series of CCD images by including realistic models of the CCD and its electronics, the telescope optics, the stellar field, the pointing uncertainty of the satellite (or Attitude Control System [ACS] jitter), and all important natural noise sources. The main questions that were addressed with this simulator were the noise properties of different photometric algorithms, the selection of the optical design, the allowable jitter amplitude, and the expected noise budget of light-curves as a function of the stellar magnitude for different parameter conditions. The results of our simulations showed that the proposed multi-telescope concept of PLATO can fulfil the defined scientific goal of measuring more than 20000 cool dwarfs brighter than m$_V$=11 with a precision better than 27 ppm/h which is essential for the study of earth-like exo-planetary systems using the transit method.
}

\maketitle

\section{Introduction}

The main objective of the PLATO mission is to study exo-planetary systems around bright stars using ultra-high precision photometry to facilitate ground-based follow-up studies. Asteroseismology of the host-stars will be the key to pinpoint characteristics such as the age of a system or the mass and density of the detected planets. A detailed description of the PLATO mission can be found in Catala (2009) and Claudi (2010).

The motivation to go to space to acquire photometry of stars for asteroseismology or detecting exo-planets is mainly the lack of atmospheric disturbances and the diurnal cycle. PLATO will be situated in a Lissajous orbit around the Lagrange point L2 which will lead to a minimization of external influences from Earth. A single field in the sky can be monitored for months to years with a very high duty cycle and with the same instrument, which leads to a very homogeneous data set, much lower photometric noise levels, and the avoidance of daily aliasing in power spectra. Nevertheless, still many noise sources remain and must be quantified in detailed studies to estimate their impact on the quality of the observations. The main unavoidable natural noise source is photon shot noise, whose relative effect on data can be minimized by collecting more photons. The instrumental noise sources which are due to the technical construction of the spacecraft and due to the limits of the electronics must be minimized to ensure that photon noise remains the major contributor of the overall noise budget. Due to the complex interplay of the noise sources, numerical simulations are ideally suited to study the overall performance of space-based observations.  

\begin{figure}
  \includegraphics[width=82mm,clip,angle=0]{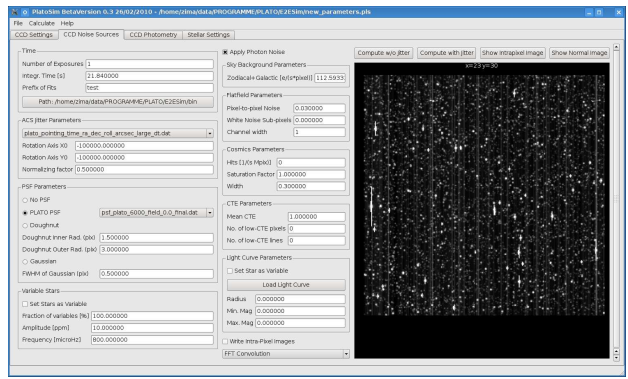}
 \caption{Screenshot of PLATOSim. On the left panels, parameters for the simulations can be set. The right panel shows a simulated image which contains several visible noise features such as saturation smearing (bright vertical spikes of stars) and frame-transfer trails (vertical lines).}
\label{fig:screenshot}
\end{figure}

PLATOSim is a software tool dedicated to the realistic simulation of CCD photometry of stars in the optical range acquired by a space-based instrument. It is designed in a way to account for the most important natural and instrumental noise sources that have an impact on the performance of the instrument. Due to the need of such a simulator for the assessment study for PLATO Mission, we created a homogeneous software tool based on pre-existing codes that had been developed for the Eddington and MONS space missions (Arentoft et al. 2004; De Ridder et al. 2006).\\ PLATOSim has been coded in C++ and features a graphical user interface (see Fig.~\ref{fig:screenshot}) based on Qt4. The contributions of the different noise sources are described through mathematical models and use a large number of parametrised inputs and numerical pre-computed input data like PSFs, jitter time-series, and sky background brightness.
 
Several improvements have been applied to the simulator compared to the pre-existing version. These concern mainly a significant increase in computational speed, a more realistic treatment of jitter, and the inclusion of two different photometric algorithms and statistical tools to estimate the noise properties of the data. The graphical interface permits in an easy way to test several different instrument set-ups by modifying the parameters for the simulations. 


\section{Modelling of CCD observations}
The modelling of the synthetic CCD image is carried out in several steps which are described in the following. The first step is to define the field-of-view (FoV) of one CCD. Important parameters here are the size of the CCD in terms of the number of pixels, the pixel-scale on the focal plane, and the positions and magnitudes of the stars. PLATO will consist of a large number of small telescopes, each with four CCDs in the focal plane. As a simplification, we only simulate the observations acquired by a single CCD and extrapolate its properties on the complete system by assuming that all telescopes behave in a similar way. Moreover, due to computational limitations, not the complete CCD but only a sub-image of typically 200x200 pixels is simulated. On such an image, all stars are assumed to have the same effective temperature and the same point spread function (PSF). Different distances from the optical axis can be simulated by using corresponding PSFs that were pre-computed for different distances. The positions of the stars on the CCD are computed assuming orthogonal projection which is a good approximation for a relatively small FoV.

We call the pixels that represent the real pixels of the CCD {\it normal} pixels. In a sub-image, each normal pixel is represented by typically up to 128x128 {\it sub-pixels}. These sub-pixels assure a realistic modelling of intra-pixel sensitivity variations of the CCD and of the ACS jitter movement which is equivalent to a few percent of a normal pixel. Only at the very end of the CCD modelling, the sub-pixels are re-binned to the normal pixel scale. After this first step of modelling, a noise-free image is created, containing the locations and the fluxes of all stars in the field-of-view.

The next step of the image simulation takes into account the physical noise sources of the CCD, the jitter movements of the spacecraft, and the instrumental optics. We model the global sensitivity variations (flat field) of the CCD by assuming a 1/f spatial power spectrum which resembles that of a typical CCD (De Ridder et al. 2006). Additionally, at sub-pixel level sensitivity variations that have white-noise characteristics are introduced. The lower sensitivity between single pixels is also modelled at sub-pixel level. 

Due to various external forces such as the solar wind and the magnetic field, and internal disturbances such as the reaction wheel noise and structural and thermal flexibility, the pointing of the satellite undergoes steady perturbations. This jitter has to be kept as small as possible due to its deteriorating effect on the observations by increasing the noise of the measured light-curves. We used a jitter time-series that has been modelled specifically for PLATO and validated using data from two space crafts, ISO and CoRoT, where pointing error history is available. The CCD image is created by summing up sub-exposures with their individual shift of the FoV, which is in the order of a few sub-pixels.

\begin{figure}
  \includegraphics[width=27mm,clip,angle=0]{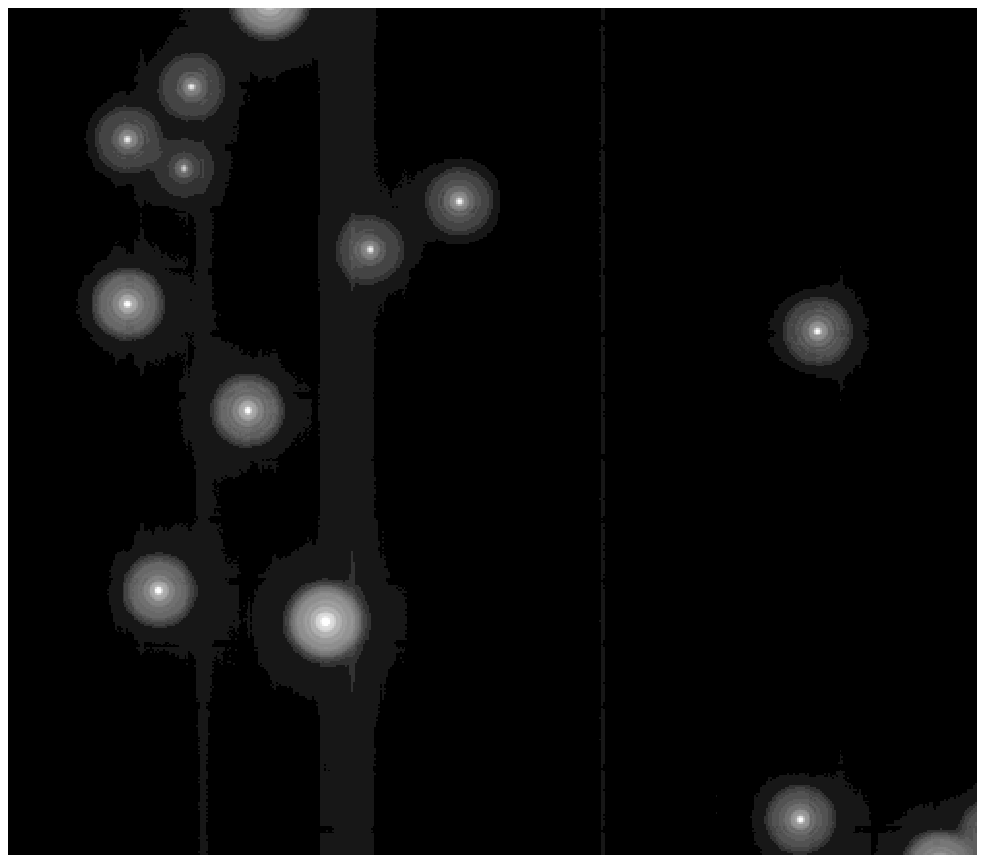}
  \includegraphics[width=27mm,clip,angle=0]{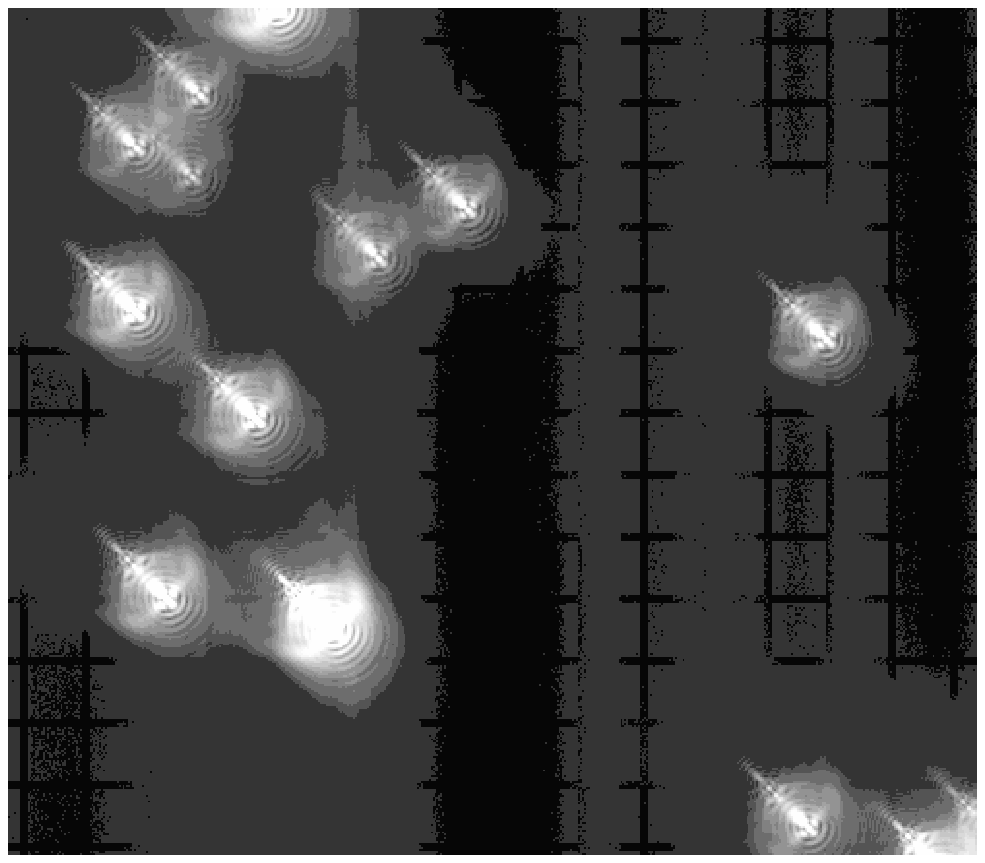}
  \includegraphics[width=27mm,clip,angle=0]{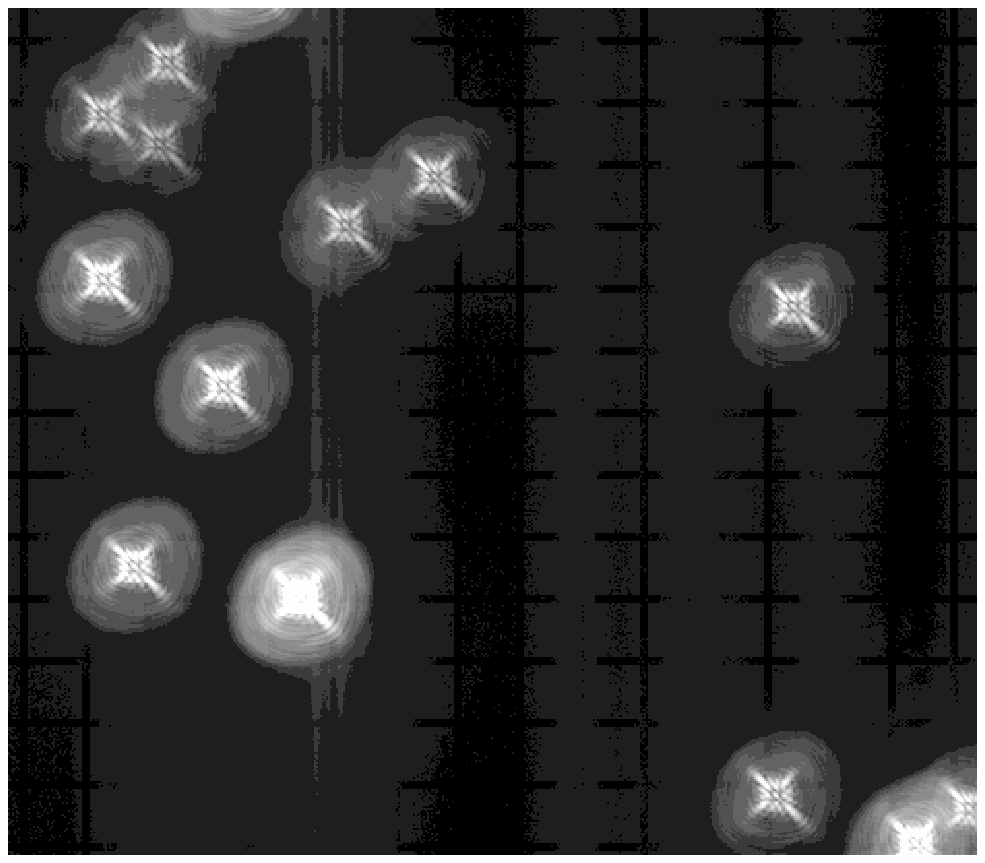}
\caption{Simulated images at sub-pixel scale of a stellar field using PSFs at different distances from the optical axis (left panel: 0$^\circ$, middle panel: 6.5$^\circ$, right panel: 14.1$^\circ$). Note the grid-like structure at the image background which arises from the lower intra-pixel sensitivity. Image generated with PLATOSim.}
\label{fig:psfs}
\end{figure}

The image is further degraded by the telescope optics which are characterised by the PSF. We used mono-chromatic PSFs that have been pre-computed for three different optical designs and different distances to the optical axis (see Fig.~\ref{fig:psfs}) for a wavelength range between 400 and 1000 nm. White light PSFs were integrated over the effective wavelength range of PLATO taking into account its wavelength dependent transmission and quantum efficiency. The resulting image is constructed by convolution in Fourier space.

The sky background of each sub-image is assumed to be uniformly lit by the zodiacal light and background objects like unresolved stars and galaxies. Through photon noise, the sky background increases the noise of the stellar photometric measurements. Values for the sky background are interpolated from tabulated data from Wehrli (1985) and Leinert et al. (1998). 

The PLATO CCDs will be working in frame-transfer mode to read out the image. This read-out mode causes the flux of all stars to be spread along the columns of the CCD and introduces an additional noise source by increasing the background flux. Although we are only simulating sub-images of the CCD with PLATOSim, for the determination of this structure the stars on the complete CCD are taken into account. As it will be done for the PLATO CCDs, we save this trailing structure and use it to correct for during data reduction. 

In the further steps, the simulated sub-image is re-binned to normal pixel scale and the photon noise is computed for each pixel. For pixels, where the flux exceeds the full well capacity (=saturation), the flux is smeared along the readout direction. Finally, charge-transfer efficiency degradation, read-out noise, and digital saturation are computed and the image is written as a FITS-file to the disk. The series of FITS files can be further analysed and reduced with standard scientific software. The  right panel of Fig.~\ref{fig:screenshot} shows the example of an image that has been simulated with PLATOSim.

We also implemented two photometric algorithms in \\PLATOSim, simple aperture and weighted mask photometry, to acquire photometric measurements of all stars on the simulated CCD in an efficient way and to compute statistics to determine their noise levels. Both algorithms take into account the sub-pixel position of the stars and the theoretical shape of the PSF to define the photometry mask which includes between 60 and 95\% of the stellar flux. We were able to demonstrate that aperture photometry produces lower noise for brighter targets where pollution due to crowding in dense fields does not pose a problem. In contrast, weighted mask photometry performs generally better for fainter stars in a crowded field due to its smaller size of the mask.

\section{Simulation study for the PLATO assessment phase}
We made an extensive set of simulations to test the performance of the photometric observations of PLATO for different set-ups of the telescope optics and the CCD in different positions in the sky. For these simulations, we used a catalogue of the proposed FoV of PLATO containing stars with m$_V \le 15$ (M. Barbieri, priv. comm.). Three different sub-fields with spatial stellar densities that account for each about a third of the FoV were studied in detail (named sparse, medium, and dense field). We computed time-series of typically 200 images and applied different photometric algorithms for all stars on the CCD and assuming that the stars have constant intrinsic magnitude. In a few simulations, low-amplitude stellar variability has been introduced to test the effect of crowding on the detectability of pulsations.

The following questions have been tackled during the assessment study of PLATO: 
\begin{itemize}
\item How many stars are affected by crowding of other sources due to confusion during photometry dependent on the number of stars per deg$^2$. Due to the large pixel size of 15 arcsec/pixel, confusion, i.e. the presence of the target star plus at least one fainter (or brighter) star in the photometry mask, is a serious problem for stars fainter than m$_V=11$. 
\item To which extent does confusion influence the detection of variable stars? To explore this question we calculated how much the photon noise of a target may be elevated due to confusion before a pulsation amplitude of 10 ppm fails to be statistically significant detectable during a 3-year run. Different scenarios, where half or all of the polluting sources pulsate with the same amplitude were explored. 
\item How do different photometric algorithms - simple aperture and weighted mask photometry - perform in terms of the noise budget? Each time-series of simulated frames has been analysed with both photometric algorithms. 
\item What is the effect of ACS jitter on the noise budget? Simulations with and without considering jitter with an rms of 0.2 arcsec have been compared.
\item Which optical design performs best? The PSFs of three different proposed optical designs and different distances to the optical axis were compared. 
\item How many stars are observable for PLATO at specific noise levels? This is essential to decide if the selected baseline design of PLATO is compliant with the mission requirements.
\end{itemize}

\begin{figure}
  \includegraphics[height=83mm,clip,angle=-90]{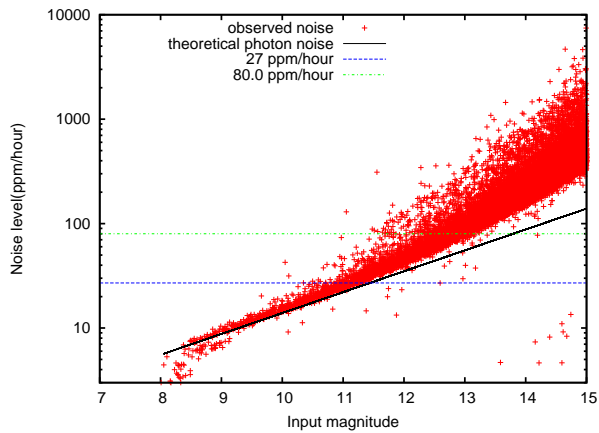}
 \caption{Expected noise level in ppm/h for observations with 40 telescopes using realistic noise modelling without jitter in the sub-field and a PSF at the optical axis. Each red cross represents the noise of a star using aperture photometry. The dark blue line is the median of the noise in bins of 0.5 mag. The theoretical photon noise is indicated as green line. Measurements that fall below this value are affected by smearing from saturated stars. The two relevant noise levels for PLATO at 27 and 80 ppm/h are marked. It can be seen that the limit for a precision of 27 ppm/h is at approximately $m_V = 11$. Towards faint magnitudes, the noise increases rapidly due to the confusion issue.}
\label{fig:noise1}
\end{figure}

\subsection{Results}
Our findings can be summarized as follows. 
\begin{itemize}
\item About 80\% of all stars with $m_V \le 12$ in the sparse and medium field, which are representative for the major part of the PLATO FoV, are not affected significantly by crowding from close-by sources.
\item Due to the smaller size of the masks, weighted mask photometry is less sensitive due to crowding than aperture photometry.
\item Stars that are affected by crowding from close-by sources also show a larger increase in noise due to ACS jitter.
\item Aperture photometry generally delivers lower noise for  $m_V \le 11$, whereas weighted mask photometry performs better for  $m_V \ge 12$ for simulations with and without jitter.
\item All three proposed optical designs perform similar with respect to crowding and the noise budget.
\item Applying photometry by using a mask derived from a slightly inaccurately modelled PSF delivers no significant signal-to-noise degradation if the used PSF deviates by a few degrees from the optical axis from the intrinsic PSF or if the stellar colour is not accurately known.
\item ACS jitter significantly increases the measured noise of the stars at all magnitudes. The increase is more pronounced in the dense field and for the near-edge PSFs. A subsequent jitter correction for photometry will be necessary to reach the required specifications of PLATO. Such an approach has been proven successful for CoRoT (De Oliveira Fialho et al. 2007, Drummond et al. 2006).
\item The number of detectable pulsating stars, assuming pulsation amplitudes of 10 ppm in a broad frequency range, decreases when more neighbouring stars add noise to the target star due to confusion and/or variability. Still, confusion is very limited except for stars in the densest fields and for $m_V \ge 12$.
\item The important magnitude range for asteroseismology between $m_V=$ 4 to 11 is only weakly affected by confusion.
\item Star counts reveal that with 40 refractive telescopes (most) scientific requirements can be fulfilled.
\begin{itemize}
\item 21000 (goal 20000) dwarfs/sub-giants at 27 ppm/h
\item 13500 (goal 1000) dwarfs/sub-giants with $m_V \le 8$ at 27 ppm/h
\item 316000 (goal 200000) dwarfs/sub-giants with $m_V \ge 8$ at 80 ppm/h
\item 52000 (goal 80000) dwarfs/sub-giants at 54 ppm/h and later 27 ppm/h

\end{itemize}
\end{itemize}

\begin{figure}
  \includegraphics[height=83mm,clip,angle=-90]{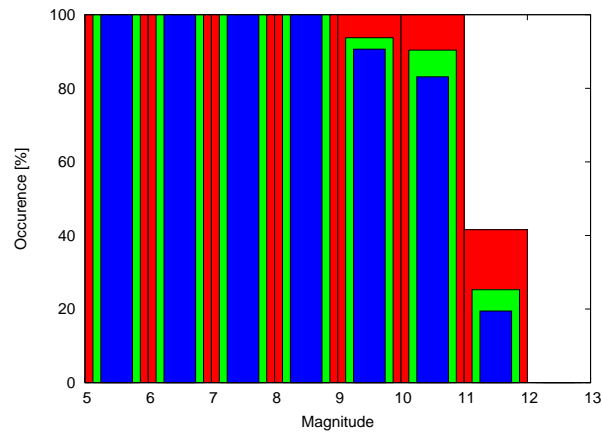}
 \caption{Results of simulations to test the detectability of a stellar oscillation signal with an amplitude of 10 ppm through confusion with neighbouring (pulsating) stars. The different coloured bars show the situation for the following assumptions (the target star is assumed to pulsate with an amplitude of 10 ppm): Red bars: Only the target star is pulsating with an amplitude of 10 ppm. The other stars in the photometry mask only contribute noise through photon noise. Green bars: 50\% of the polluting stellar sources in the photometry mask pulsate with an amplitude of 10 ppm in a broad frequency range. Blue bars: All polluting stellar sources in the photometry mask pulsate with an amplitude of 10 ppm in a broad frequency range. Above $m_V=12$, no pulsations with an amplitude of 10 ppm can be detected. These simulations were done for a medium dense field and a PSF at 6$^\circ$ from the optical axis. }
\label{fig:noise1}
\end{figure}

\section{Future outlook}
The ESA Science Program Committee has decided in its meeting on Feb. 17/18, 2010 that the PLATO mission proceeds to a Phase B definition study which will end in June, 2011.  A modification of the current design of the mission, to enable compliance with the mission requirements regarding especially the mass budget and schedule is necessary. Therefore, in the new design for PLATO, the number of telescopes will be reduced, the telescope optics will be modified (fewer lenses), the FoV will be increased and the observation of brighter stars is anticipated. These modifications will require more simulations to fine-tune the mission design and ensure that the scientific requirements will be met.

\acknowledgements
The research leading to these results has received funding from the European Research Council under the European Community's Seventh Framework Programme (FP7/2007--2013)/ERC grant agreement n$^\circ$227224 (PROSPERITY), as well as from the Research Council of K.U. Leuven grant agreement GOA/2008/04. WZ thanks R\'eza Samadi for fruitful discussions and important scientific input to this work.


\begin{thebibliography}{}
\bibitem{} Arentoft, T., Kjeldsen, H., De Ridder, J., Stello, D.: 2004, ESASP 538, 59 
\bibitem{} Catala, C.: 2009, CoAst 158, 330
\bibitem{} Claudi, R.: 2010, Ap\&SS 57
\bibitem{} De Oliveira Fialho, F., Lapeyrere, V., Auvergne, M., Drummond, R., Vandenbussche, B., Aerts, C., Kuschnig, R., Matthews, J. M.: 2007, PASP 119, 337
\bibitem{} De Ridder, J., Arentoft, T., Kjeldsen, H.: 2006, MNRAS 365, 595
\bibitem{} Drummond, R., Vandenbussche, B., Aerts, C., De Oliveira Fialho, F., Auvergne, M.: 2006, PASP 118, 874
\bibitem{} Leinert, C., Bowyer, S., Haikala, L. K., Hanner, M. S. et al.: 1998, A\&AS, 127, 1
\bibitem{} Wehrli, C.: 1985, Extraterrestrial Solar Spectrum, in Pub. No. 615 of the Physikalisch-Meteorologisches Observatorium + World Radiation Center (PMO/WRC) Davos Dorf, Switzerland

\end{thebibliography}
\end{document}